\documentstyle[prl,aps,twocolumn]{revtex}
\newcommand{\bacugeo}{Ba$_2$CuGe$_2$O$_7$ }

\begin{document}

\draft

\title{Experimental evidence for Shekhtman-Entin-Wohlman-Aharony (SEA) interactions in \bacugeo}

\author{A. Zheludev$^{(1)}$ \and S. Maslov$^{(1)}$ \and I. Tsukada $^{(2)}$
\and I. Zaliznyak $^{(3,4)}$ \and L. P. Regnault $^{(5)}$
\and T. Masuda$^{(2)}$ \and K. Uchinokura$^{(2)}$
\and R.~Erwin $^{(3)}$ \and G. Shirane$^{(1)}$}

\address{(1) Brookhaven National  Laboratory,
Upton, NY 11973-5000, USA. (2) Department of Applied Physics, The
University of Tokyo, 7-3-1 Hongo, Bunkyo-ku, Tokyo 113-8656, Japan. (3)
NIST Center for Neutron Research, National Institute of Standards and
Technology, MD 20899, USA. (4) Department of Physics and Astronomy,
Johns Hopkins University, MD 21218 USA, and P.~L.~Kapitza Institute for
Physical Problems, Moscow, Russia. (5) RFMC/SPSMS/MDN, CENG, 17 rue des
Martyrs, 38054 Grenoble Cedex, France.}

\date{\today}

\maketitle

\begin{abstract}
New neutron diffraction and inelastic neutron scattering experiments on
\bacugeo suggest that the previously suggested model for the magnetism
of this material (an ideal sinusoidal spin spiral, stabilized by
isotropic exchange and Dzyaloshinskii-Moriya interactions) needs to be
refined. Both new and previously published experimental results can be
quantitatively explained by taking into account the
Shekhtman-Entin-Wohlman-Aharony (SEA) term, a special anisotropy term
that was predicted to always accompany Dzyaloshinskii-Moriya
interactions in insulators. Our experimental results present the first
clear evidence to that SEA interactions can lead to substantial
observable effects in a real magnetic system.
\end{abstract}

Among the more exotic magnetic interactions in solids is the so-called
asymmetric exchange, first predicted theoretically by Dzyaloshinskii
\cite{Dzyaloshinskii57}. Unlike conventional Heisenberg exchange
coupling, that is proportional to the {\it scalar} product
$\bbox{S}_{1}\cdot
\bbox{S}_{2}$ of interacting spins, asymmetric exchange is proportional
to the corresponding {\it vector} product. In the spin Hamiltonian it is
usually written as $\bbox{D} (\bbox{S}_{1}\times \bbox{S}_{2})$, where
$\bbox{D}$ is the Dzyaloshinskii vector associated with the bond between
the two interacting magnetic ions. A microscopic model for asymmentric
exchange interactions was first proposed by Moriya \cite{Moriya60}, and
is essentially an extension of the Anderson superexchange
mechanism\cite{Anderson59}, that allows for spin-flip hopping of
electrons. While forbidden by symmetry in centrosymmetric crystal
structures, Dzyaloshinskii-Moriya (DM) interactions were found to be
active in a number of non-centric compounds, where they lead to either a
weak ferromagnetic- or helimagnetic- distortion of the collinear
magnetic state \cite{Ishikawa-Shirane77,Ishikawa84,Lebech89}. The
inclusion of the DM term breaks $O(3)$ invariance of the originally
isotropic Heisenberg spin Hamiltonian, reducing the symmetry to $O(2)$:
to take full advantage of the cross product term the interacting spins
must be perpendicular to the vector $\bbox{D}$. DM interactions thus
play a role of an effective two-ion easy-plane anisotropy, with the easy
plane normal to the vector $\bbox{D}$.

Only recently Schekhtman, Entin-Wohlman and Aharony realized that there
is {\it more} to Moriya's mechanism than just the vector-product term
\cite{Shekhtman92,Shekhtman93}. For very fundamental reasons the DM
cross-product must always be accompanied by a two-ion easy axis
anisotropy term that exactly compensates the easy-plane effect of the
vector product. The additional (SEA) term can to a good approximation be
written as $\case{1}{2J}(\bbox{S}_1\bbox{D})(\bbox{S}_2\bbox{D})$, where
$J$ is the Heisenberg (isotropic) component of superexchange coupling.
Often referred to as ``hidden symmetry'', the SEA term {\it restores}
$O(3)$ invariance of the Hamiltonian, at least locally. Originally, the
SEA term was invoked to explain the spin anisotropy in the orthorhombic
phase of La$_2$CuO$_4$ \cite{Shekhtman92,Shekhtman93,Entin94}. It was
later realized that this term alone can not account for all the observed
effects, particularly for the magnetic anisotropy seen in the tetragonal
phase \cite{Koshibae93,Yildirim94,Yildirim95}. To our knowledge, to date
there has been no ``clean'' experimental evidence unambiguously pointing
to the presence of SEA interactions. In the present paper we present
such experimental data for the helimagnetic insulator
Ba$_2$CuGe$_2$O$_7$. We demonstrate that only by taking into account the
SEA term one can obtain qualitatively and quantitatively correct
predictions for the magnetic structure and spin wave spectrum.

As was shown in a series of recent publications
\cite{Zheludev96BACUGEO,ZM97BACUGEO-B,ZM97BACUGEO-L,ZM98BACUGEO},
\bacugeo is a particularly useful model system for studying DM
interactions. The magnetism of this compound is due to Cu$^{2+}$ ions
that form a square lattice in the $(a,b)$ tetragonal plane of the
crystal. The principal axes of this square lattice, hereafter referred
to as the $x$ and $y$ axis, run along the $[1 1 0]$ and $[1
\overline{1} 0]$ crystallographic directions, respectively. To complete the coordinate
system we shall choose the $z$ along the $[0 0 1]$ direction. In the
magnetically ordered phase (below $T_{N}=3.2$~K) all spins lie in the
$(1\overline{1}0)$ plane. The magnetic propagation vector is
$(1+\zeta,\zeta,0)$, where $\zeta=0.0273$, and $(1,0,0)$ is the
antiferromagnetic zone-center. The magnetic structure is a distortion of
a N\'{e}el spin arrangement: a translation along
$(\case{1}{2},\case{1}{2},0)$ induces a spin rotation by an angle
$\phi={2 \pi}{\zeta}\approx9.8^{\circ}$ (relative to an exact
antiparallel alignment) in the $(1,-1,0)$ plane. Along the $[1
\overline{1} 0]$ direction nearest-neighbor spins are perfectly
antiparallel. Spins in adjacent Cu-planes are aligned parallel to each
other. Only nearest-neighbor in-plane isotropic superexchange
antiferromagnetic interactions are important ($J\approx 0.96$~meV, as
determined by the measured spin wave bandwith \cite{Zheludev96BACUGEO}).
The helical state is stabilized by DM interactions. In the current model
Dzyaloshinskii vectors for nearest-neighbor Cu-Cu pairs lie in the
$(x,y)$ plane and are oriented perpendicular to their corresponding
bonds: $\bbox{D}||y$ for an $x$-bond and $\bbox{D}||x$ for $y$-bonds,
respectively (Fig.~1 in Ref.~\cite{ZM97BACUGEO-L}). The corresponding
energy scale is $D \approx 0.17$~meV. In the discussion below we shall
use the numerical values for $J$ and $\zeta$ quoted above as given, and
perform all calculations without using any adjustable parameters.

The magnetic structure of \bacugeo is expected to be particularly
sensitive to the presence of SEA interactions. Indeed, in \bacugeo SEA
easy axes that correspond to the $y$- Cu-Cu bonds are parallel to the
$x$ axis, i.e., are {\it in the plane of spin rotation}. Regions of the
slowly rotating spin spiral where the local staggered magnetization
$\bbox{l}$ is almost parallel to $x$ become more energetically favorable
than those where $\bbox{l}$ is almost parallel to $z$ (crystallographic
$c$-axis). SEA anisotropy must therefore lead to a distortion of the
ideal sinusoidal spiral, and modify the period of the structure. The SEA
term is expected to produce exactly the same distortion as a magnetic
field $H$ applied along the $z$ axis: the latter also has the effect of
forcing the local staggered magnetization into the $(x,y)$ plane. The
role of a $z$-axis field is rather dramatic and has been studied in
detail \cite{ZM97BACUGEO-L,ZM98BACUGEO}. The period of the spiral
increases with increasing $H$ and diverges at $H_c\approx 2.15$~T,
resulting in a commensurate spin-flop antiferromagnetic state at
$H>H_c$. For $0<H<H_c$ the spin structure is described as a ``soliton
lattice'' where regions of the commensurate phase are interrupted by
regularly spaced antiferromagnetic domain walls. In the soliton phase,
in addition to the principal magnetic Bragg peaks at
$(1\pm\zeta,\pm\zeta, 0)$, characteristic of an ideal spiral, one
expects to see all odd magnetic Bragg harmonics at $(1\pm 3\zeta,
\pm 3\zeta, 0)$, $(1\pm 5\zeta, \pm 5\zeta, 0)$, etc. By comparing the
experimental field dependencies of $\zeta$ and the higher-order Bragg
peaks to theoretical predictions for the ``DM-only''
(Ref.~\cite{ZM97BACUGEO-L,ZM98BACUGEO}) and ``DM+SEA'' models we can
hope to obtain direct evidence for SEA interactions in
Ba$_2$CuGe$_2$O$_7$.

We can make the above discussion quantitative by including the SEA term
into the phenomenological energy functional that was previously used to
describe the behaviour of \bacugeo in the framework of the DM-only model
\cite{ZM97BACUGEO-L,ZM98BACUGEO}. This procedure is rather
straightforward and the principal conclusion is that all previously
obtained DM-only results can be recycled, by replacing $H$ in all
equations by the {\it effective} field
\begin{equation}
H^{\rm eff}=\sqrt{H^2+2A\rho_S
/(\chi_{\perp}-\chi_{\|})}.\label{heff}
\end{equation}
 Here $\rho_S\approx JS^2$ is the spin
stiffness, $\chi_{\perp}$ and $\chi_{\|}$ are the local transverse and
logitudinal susceptibilities, respectively, and the SEA term is
represented by $A=\alpha^2/2\approx D^2/2J^2$. The parameter $\alpha$ is
defined by $\tan \alpha \equiv D/J$, and is equal to the spin rotation
angle $\phi$ in the DM-only model. According to our continuous-limit
calculations, in the DM+SEA model $\alpha
\equiv \arctan(D/J)=\case{32}{31}\phi$.

First, let us consider the field dependence of the incommensurability
parameter $\zeta$ that for the DM+SEA model can be obtained by replacing
$H$ by $H^{\rm eff}$ in Eqs. 4 and 7 in Ref.~\cite{ZM97BACUGEO-L}:
\begin{eqnarray}
\frac{\zeta(H)}{\zeta(0)}=\frac{\pi^2}{4E(\beta)K(\beta)}\\
\frac{H^{\rm eff}}{H_c^{\rm eff}}=\frac{\beta}{E(\beta)}
\end{eqnarray}
Here $\beta$ is an implicit variable and $K$ and $E$ are complete
elliptic integrals of the first and second kind, respectively. In
Fig.~\ref{zeta} we replot the $\zeta(H)$ data from
Ref.~\cite{ZM97BACUGEO-L} in reduced coordinates. The solid and dashed
lines are the theoretical curves plotted with and without taking into
account SEA interactions, respectively. We see that the inclusion of the
SEA term hardly affects the {\it shape} of the $\zeta(H)$ curve.
However, the theoretical prediction for $H_c$ is substantially different
in the DM-only and DM+SEA models. Combining Eq.~\ref{heff} from above
with Eq~5 in Ref.~\cite{ZM97BACUGEO-L} one readily obtains:
\begin{equation}
H_c=\alpha {\sqrt{\pi^2-4}\over 2} \sqrt{\rho_s \over
\chi_{\perp}-\chi_{\|}}, \label{hc}
\end{equation}
For the low-temperature limit in \bacugeo we can use the classical
expressions $\rho_s=JS^2=0.24$ meV, $\chi_{\|}=0$ and $\chi_{\perp}=(g_c
\mu_B)^2/8J$, where $g_c=2.47$ is the $c$-axis gyromagnetic ratio for Cu$^{2+}$
in \bacugeo \cite{Sasago-ESR}. One can expect these classical estimates
to be rather accurate, as they rely on the {\it effective} exchange
constant $J$, that itself was determined from fitting the {\it
classical} spin wave dispersion relations to inelastic neutron
scattering data \cite{Zheludev96BACUGEO}. Substituting
$\alpha=2\pi\case{32}{31}\zeta=0.177$, we get the estimate for the
critical field $H_c=2.05$~T. This value is much closer to the
experimental value $H_c\approx2.15$~T, than our previous estimate
$H_c\approx2.6$~T \cite{hcwrong}, obtained without taking into account
the SEA term.

As mentioned, SEA interactions have a substantial influence on the
intensity of higher-order Bragg harmonics. In the DM-only model in zero
field higher-order Bragg reflections are totally absent. For the DM+SEA
model, combining our expression for $H^{\rm eff}$ with Eqs.~17,18 in
Ref.~\cite{ZM98BACUGEO}, for the relative intensities of the 1st and 3rd
harmonics, in the small field limit (weakly distorted spiral)
$|\phi-\alpha|\ll
\phi$ we get:
\begin{equation}
{I_3 \over I_1}=
\left[{1 \over 16} + \left({\pi^2 \over 64}-{1 \over 16}
\right) \left({H \over H_c}\right)^2 \right]^2\label{i3}
\end{equation}
In zero field this gives $I_3/I_1=1/256\approx 4\times 10^{-3}$. To
verify the relation (\ref{i3}) we performed new magnetic neutron
scattering experiments on \bacugeo single crystal samples. The
measurements were done in two experimental runs, on the IN-14 3-axis
spectrometer at the Institut Laue Langevein (ILL) in Grenoble, and the
SPINS spectrometer at the Cold Neutron Research Facility at the National
Institute of Standards and Technology (NIST). The samples were similar
to those used in previous studies\cite{ZM98BACUGEO}. In each experiment
the crystals were mounted with their $c$-axes vertical, making $(h,k,0)$
wave vectors accessible for measurements. The magnetic field was
produced by standard split-coil superconducting magnets. The sample
environment was either a pumped-$^4$He or pumped-$^3$He cryostat. The
data were collected at temperatures in the range range 0.35--5~K.
Neutrons of energies 3.5~meV or 2.5~meV were used in most cases. A
$40'-S-40'-A-40'$ collimation setup was used with a Be filter positioned
in front of the sample. In Fig. \ref{diff}(a,b) we show some typical
elastic scans along the $(1+\epsilon, \epsilon,0)$ reciprocal-space line
measured in
\bacugeo at low temperatures in zero and $H=1$~T applied fields. Even in the zero-field data
in addition to the 1st-order principal magnetic reflection one clearly
sees the 3rd order harmonic. The measured field dependence of $I_3/I_1$
(ratio of $Q$-integrated intensities) is shown in Fig. \ref{diff}(c). In
our measurements we have taken special care to verify that the relative
intensities of the two peaks are totally independent of the $T-H$
history of the sample. The solid and dashed lines in Fig. \ref{diff}(c)
represent the predictions of the DM+SEA (Eq.~\ref{i3}) and DM-only
(Ref.~\cite{ZM97BACUGEO-L,ZM98BACUGEO}) models, respectively. For these
theoretical curves we used the same numerical values as in our estimates
for $H_c$, and no adjustable parameters. An almost perfect agreement
between the DM+SEA model and the experimental data is apparent, and so
is the failure of the DM-only model.

It is clear that the SEA anisotropy term, in addition to modifying the
ground state, will affect the spin wave spectrum. For an ideal spin
spiral (DM-only model) the classical spin wave dispersion relations can
easily be obtained analytically using the Holstein-Primakov formalism.
The result of such a calculation for \bacugeo is shown in
Fig.~\ref{disp}(a). Two acoustic branches (hereafter referred to as the
$\pm \zeta$ modes) emerge from the two magnetic Bragg peaks at $(1\pm
\zeta, \zeta, 0)$. A third branch (the 0-mode) has a gap at the antiferromagnetic
zone-center, equal to the Dzyaloshinskii parameter $D$. This branch
almost excatly passes through the intersection point of the $\pm\zeta$
modes. The actual dispersion curves in \bacugeo were measured in
constant-$Q$ inelastic scans using the experimental setups described
above in the fixed-incident-energy mode. In these experiments the
typical energy resolution, as measured by scanning through incoherent
scattering, was $0.075$~meV FWHM at zero energy transfer. A strong
incoherent signal and the ``tails'' of magnetic bragg reflections
prevented us from collecting reliable data for energy transfers of less
than $\approx 0.17$~meV. A typical inelastic scan (raw data) is shown in
Fig.~\ref{exdata}. Combined data from the two series of experiments are
summarized in the experimental dispersion relations in
Fig.~\ref{disp}(b) (symbols). Two prominent features of the measured
dispersion curves are to be discussed here. i) The two $\pm
\zeta$ modes do not intersect at the N\'{e}el point. Instead, at
$\bbox{Q}=(1,0,0)$ there is a clear repulsion between these two
branches. This repulsion is again manifest at
$\bbox{Q}=(1+2\zeta,2\zeta,0)$ and is seen as a discontinuity in the
$+\zeta$- branch. ii) At the antiferromagnetic zone-center the 0-mode
lies visibly lower than the extrapolated point of intersection of the
$\pm\zeta$ branches (dashed lines). Obviously, the DM-only model fails
to reproduce the observed dispersion relations. An exact analytical
calculation of the spin wave spectrum in the presence of SEA
interactions is not possible. However, in the limit
$|\phi-\alpha|\ll\phi$, a condition well satisfied in \bacugeo in zero
applied field, one can use the series expansion method described in
Refs.~\cite{Zhitomirsky96,Zaliznyak95}, and leaving only the
lowest-order terms in $A$. Performing this somewhat tedious calculation
for Ba$_2$CuGe$_2$O$_7$, we obtain the dispersion relations shown in
solid lines in Fig.~\ref{disp}(b), and find very good agreement with
experiment with {\it no adjustable parameters}.

In summary, both the static and dynamic magnetic properties of
\bacugeo contain {\it quantitative} evidence of the SEA anisotropy term.
The issue that we tried to address above is not whether or not SEA
interactions {\it exist}: if one believe Anderson's and Moriya's
superexchange mechanisms, one is forced to accept the presence of the
SEA term as well. Rather, our results for the first time show that SEA
interactions can result in very interesting measurable effects in a real
magnetic system. We also would like to emphasize that the new results do
not undermine our previous conclusions regarding the
commensurate-incommensurate Dzyaloshinskii transition in
Ba$_2$CuGe$_2$O$_7$. The transition occurs exactly as described, only
now we have a more accurate theoretical estimate for the critical field
$H_c$.

This study was supported in part by NEDO (New Energy and Industrial
Technology Development Organization) International Joint Research Grant
and the U.S. -Japan Cooperative Program on Neutron Scattering. Work at
BNL was carried out under Contract No. DE-AC02-76CH00016, Division of
Material Science, U.S. Department of Energy. Experiments at NIST were
partially supported by the NSF under contract No. DMR-9413101.

\begin{figure}
\caption{Field dependence of the incommensurability parameter $\zeta$, as
previously measured in
\bacugeo (Refs. \protect{\cite{ZM97BACUGEO-L,ZM98BACUGEO}}). The
solid and dashed lines are theoretical predictions that do (this work)
or do not (Ref. \protect{\cite{ZM97BACUGEO-L}}) take into account SEA
interactions, respectively.}
\label{zeta}
\end{figure}

\begin{figure}
\caption{Typical elastic scans along the $(1,1,0)$ direction in the vicinity of
the antiferromagnetic zone-center $(1,0,0)$, measured in
\bacugeo at $T=0.35$~K in zero field (a) and in a $H=1.6$~T magnetic field applied
along the $c$-axis (b). (c)The square root of the measured ratio of the
intensities of the $(1+3\zeta,3\zeta,0)$ and $(1+\zeta,\zeta,0)$ peaks
plotted against the square of the applied field. The lines are guides
for the eye in (a) and (b) and theoretical curves in (c), as in
Fig.~\protect{\ref{zeta}}.}
\label{diff}
\end{figure}

\begin{figure}
\caption{(a) Classical spin wave dispersion relations calculated for \bacugeo without
taking into account SEA interactions. (b) Solid lines: same, with the
SEA term included. Symbols: experimental dispersion curves measure in
\bacugeo at $T=0.35$~K and $T=1.5$~K with inelastic neutron scattering.}
\label{disp}
\end{figure}

\begin{figure}
\caption{ Typical inelastic scans measured in \bacugeo at $T=1.5$~K and $T=3$~K (combined data in
lower plot). The shaded curves represent the individual Gaussians in a
multi-peak fit (heavy solid line). The gray area shows the position of a
``Bragg-tail'' spurious peak.}
\label{exdata}
\end{figure}


\begin{thebibliography}{10}
\bibitem{hcwrong}
The numerical estimate for the critical field in Ba$_2$CuGe$_2$O$_7$ (DM-only
  model), quoted in Refs.\protect\cite{ZM97BACUGEO-L,ZM98BACUGEO} ($H_c=3.3$~T)
  is incorrect. Substituting the classical parameters listed in the end of
  Ref.\protect\cite{ZM97BACUGEO-L} into the expression for $H_c$ in the same
  paper, one actually gets $H_c=2.6$~T.
\end{thebibliography}
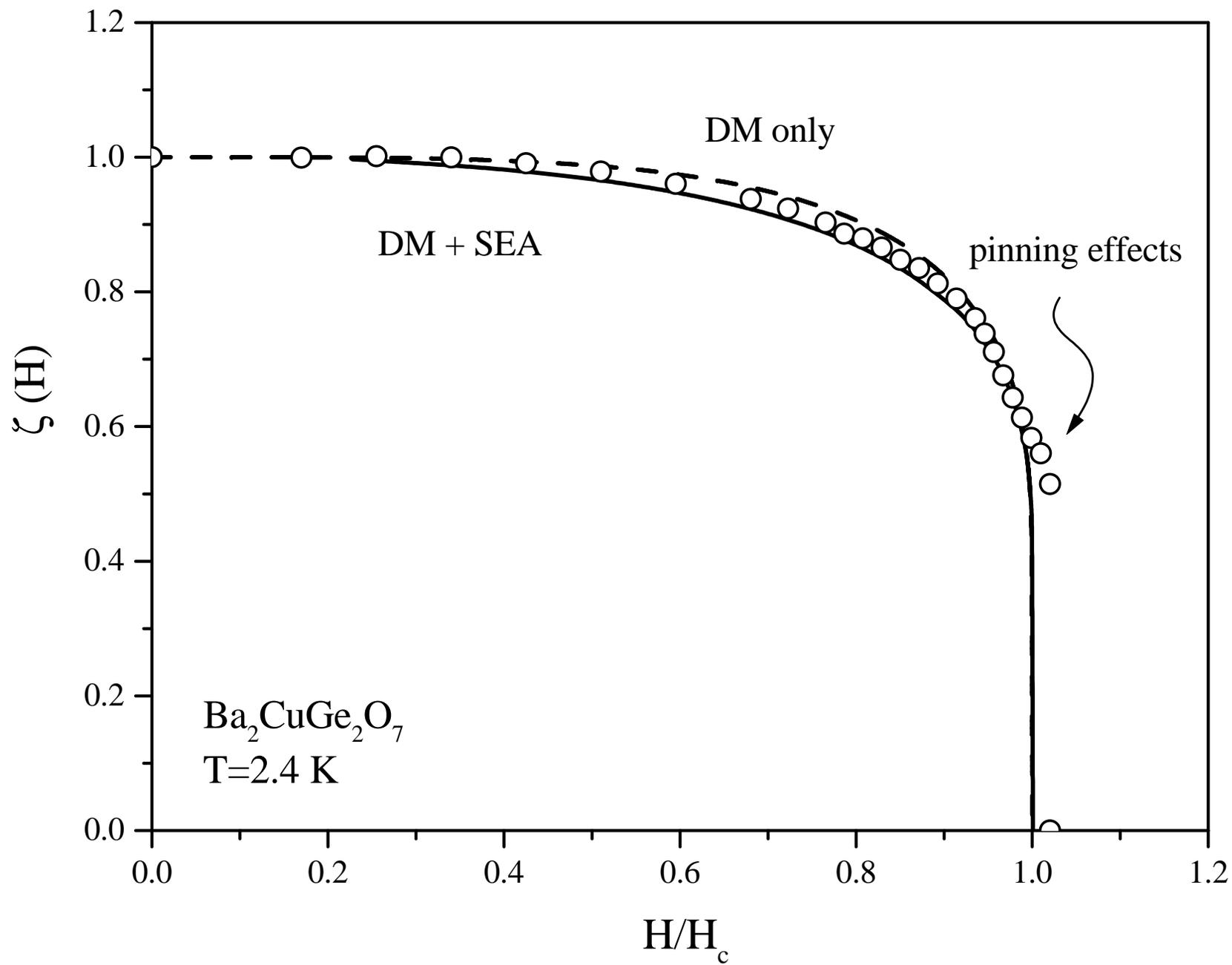
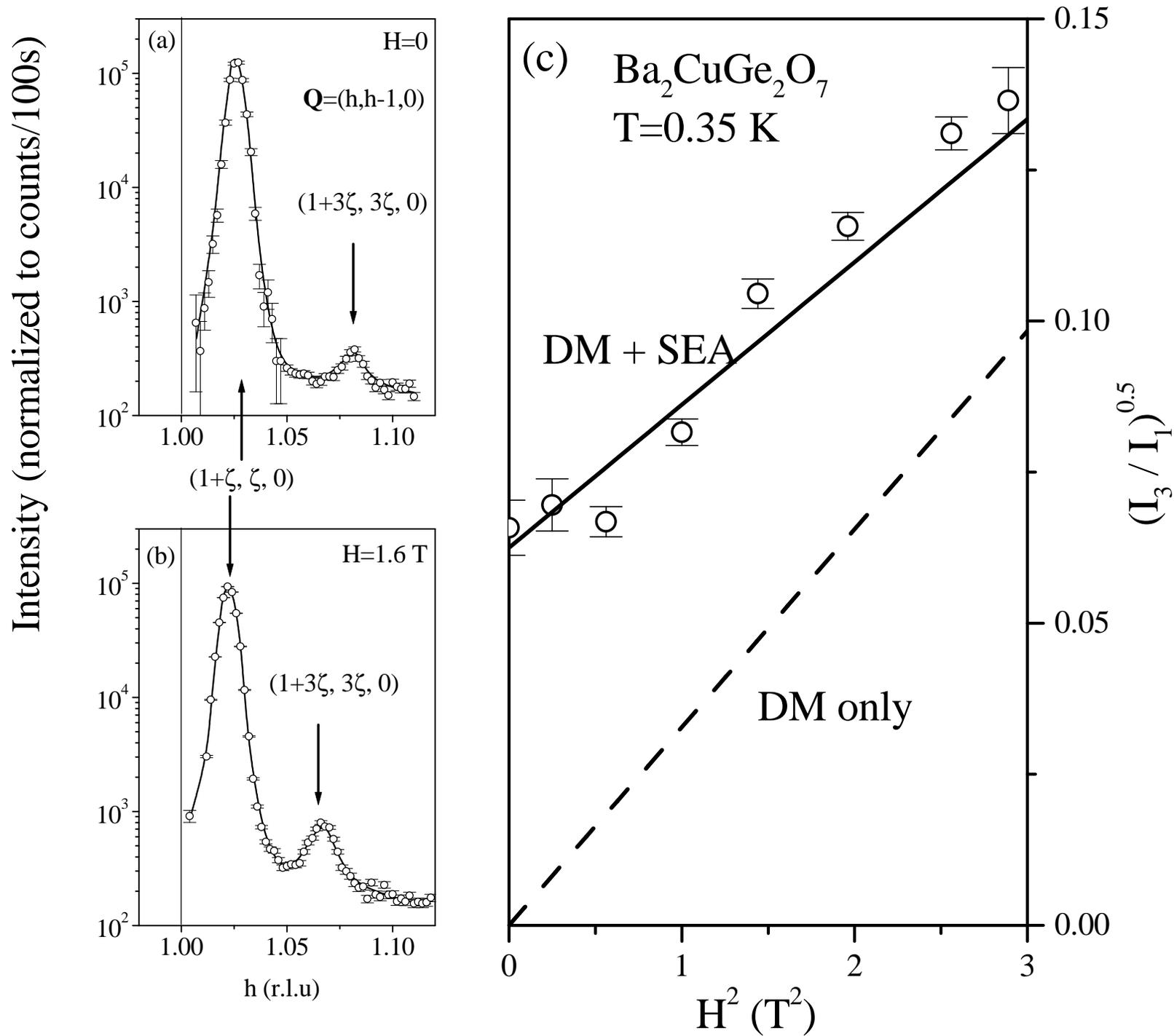
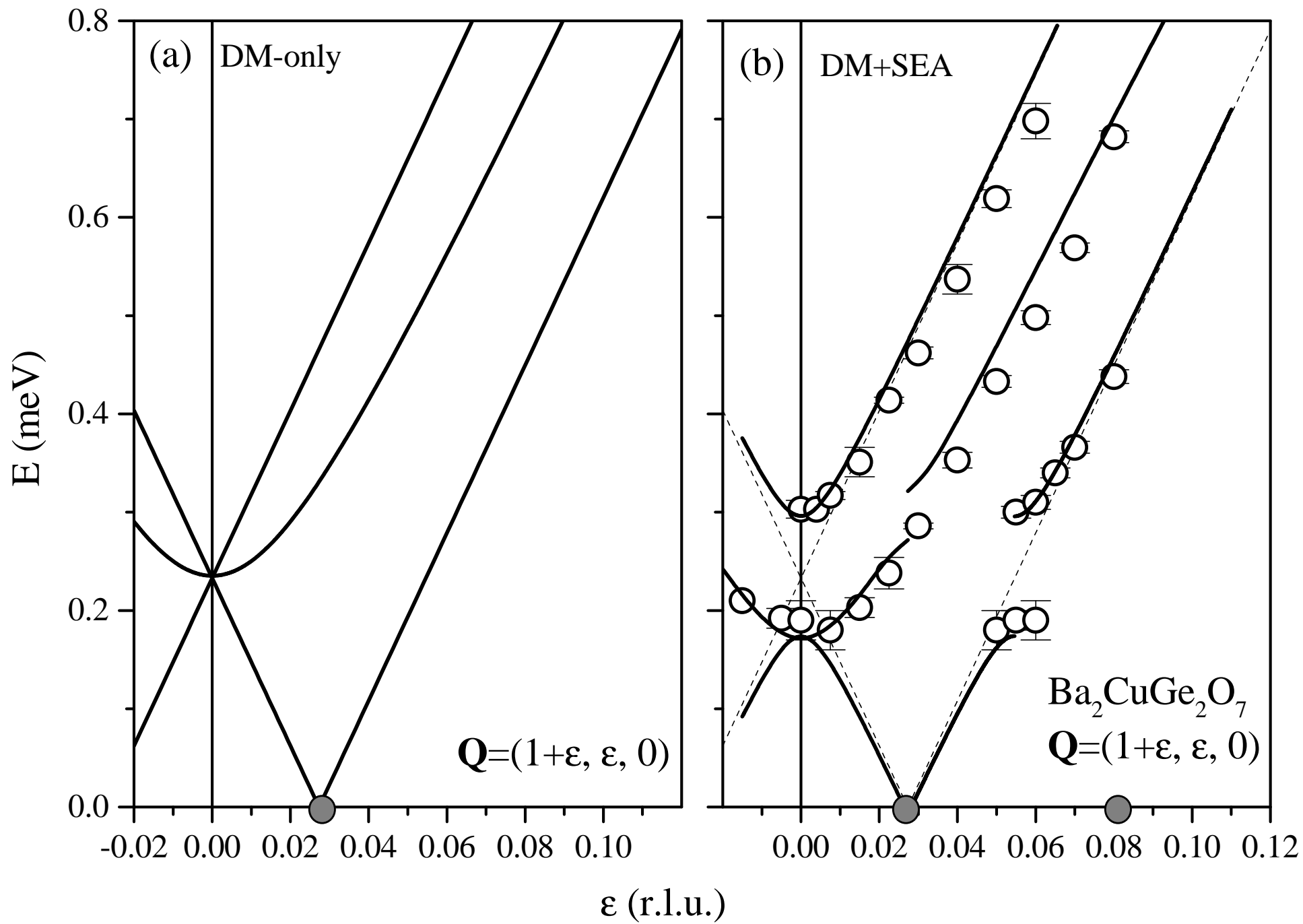
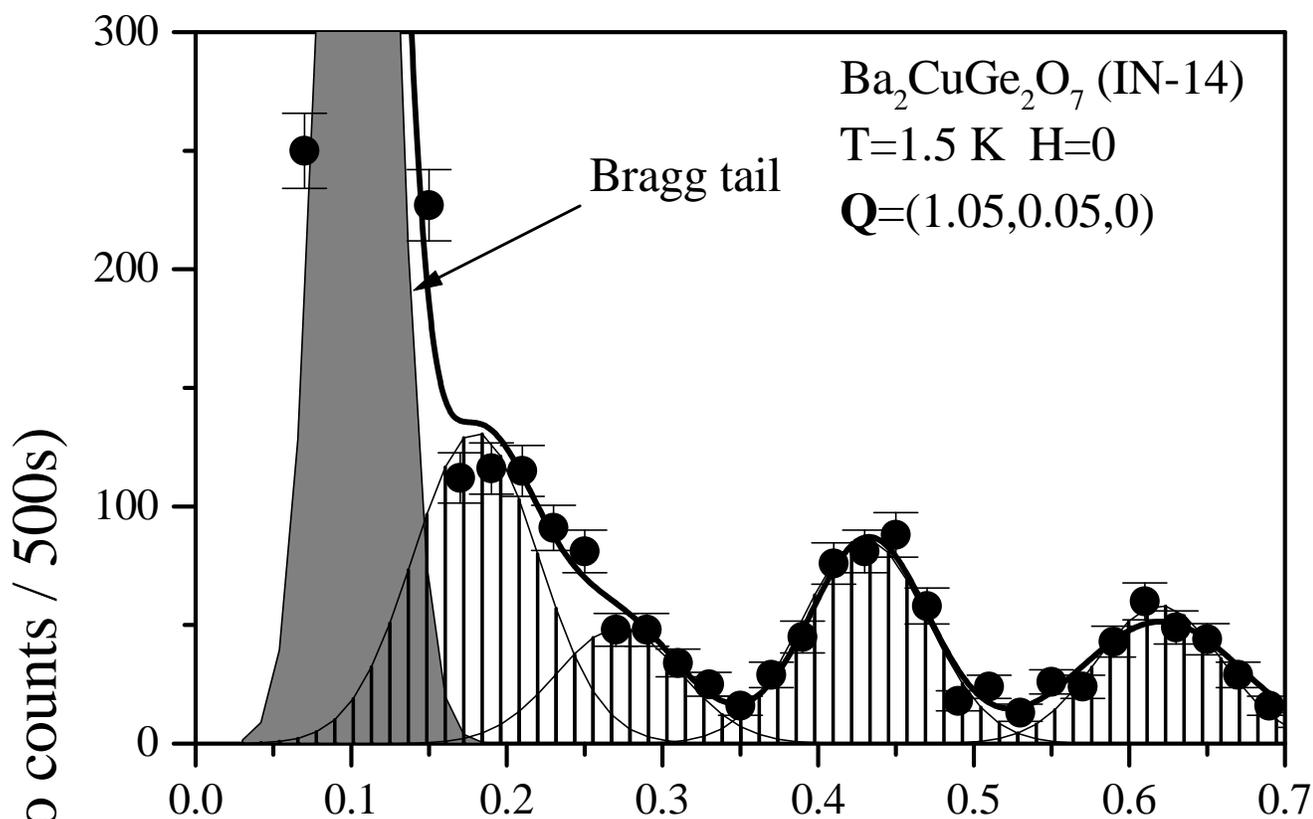
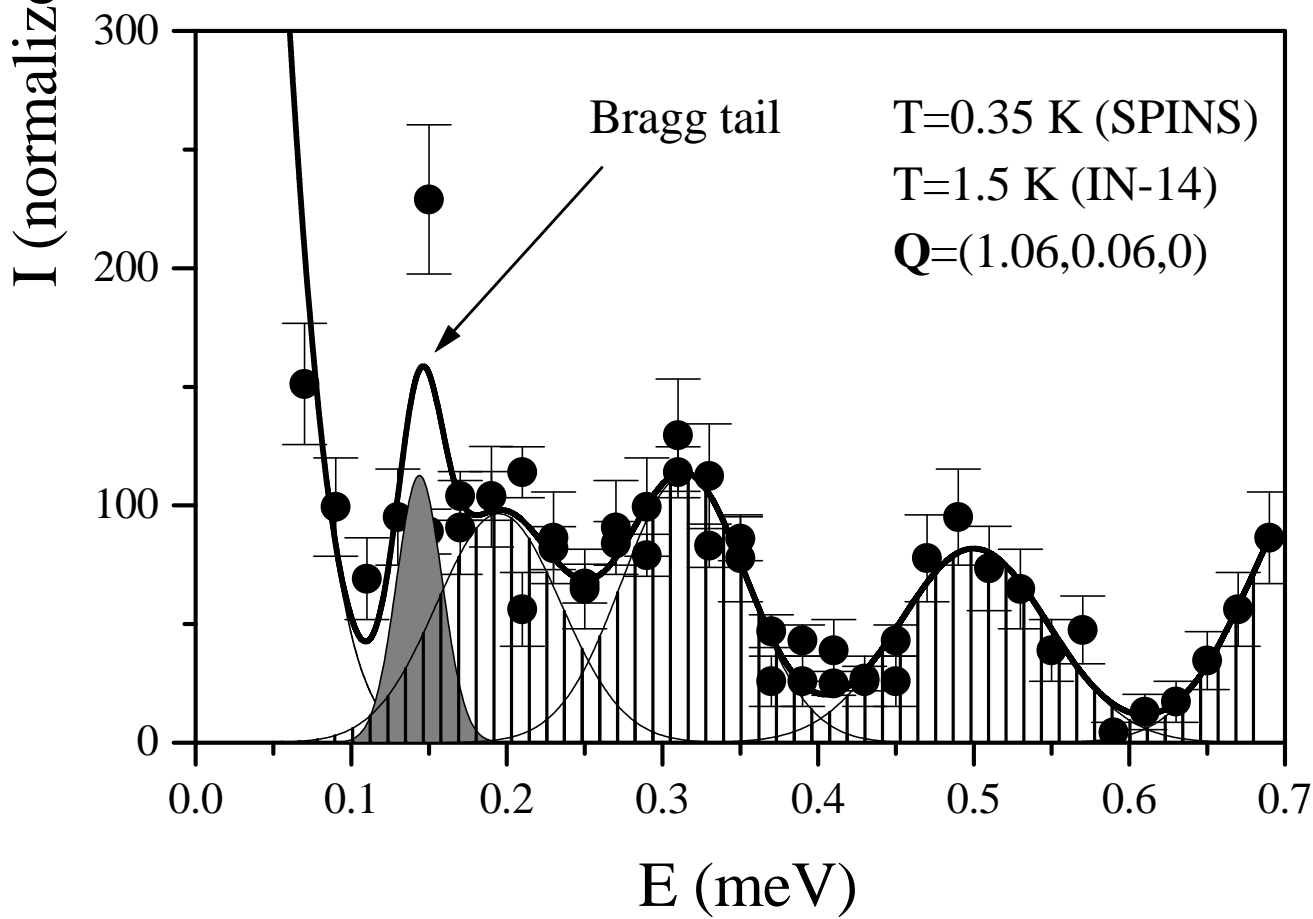
\end{document}